\documentclass[traditabstract]{aa}

\RequirePackage{lineno}

\usepackage{graphicx}
\usepackage{txfonts}
\usepackage{lineno} 
\usepackage{natbib}
\usepackage{ulem}
\usepackage{setspace} 
\bibpunct{(}{)}{;}{a}{}{,}

\newcommand{\cmc}{\mbox{$\mbox{cm}^{-3}$}}
\newcommand{\cmd}{\mbox{$\mbox{cm}^{-2}$}}
\newcommand{\kms}{\mbox{km$\,$s$^{-1}$}}
 
\newcommand{\gcm}{\mbox{$\mbox{g} \, \mbox{cm}^{2}$}} 
\newcommand{\lsun}{\mbox{$L_\odot$}}
\newcommand{\msun}{\mbox{$M_\odot$}}

\newcommand{\nhtwo}{\mbox{$N_{\mbox{\tiny H$_2$}}$}}

\newcommand{\micron}{\mbox{$\mu$m}}

\newcommand{\lbol}{\mbox{$L_{\mbox{\tiny bol}}$}}
\newcommand{\lsubmm}{\mbox{$L_{\mbox{\tiny $>$$350\mu$m}}$}}
\newcommand{\hii}{H \mbox{\scriptsize II }}

\newcommand{\tdust}{\mbox{$T_{\mbox{\tiny dust}}$}}

\begin{document}

   \title{ The \textit{Herschel}\thanks{
   Herschel is an ESA space observatory with science instruments
provided by European-led Principal Investigator consortia and with important participation from NASA. {See \it http://herschel.esac.esa.int/}}
view of massive star formation in G035.39--00.33:\\
Dense and cold filament of W48 undergoing a mini-starburst\thanks{
   Figures 9--11 are only available in electronic form at http://www.aanda.org}}
  \author{Q.~Nguy$\tilde{\hat{\rm e}}$n~Lu{\hskip-0.65mm\small'{}\hskip-0.5mm}o{\hskip-0.65mm\small'{}\hskip-0.5mm}ng\inst{1},
      F.~Motte\inst{1},
      M.~Hennemann\inst{1},
      T.~Hill\inst{1}, 
      K.~L.~J.~Rygl\inst{2},
      N.~Schneider\inst{1},
      S.~Bontemps\inst{3},
      A.~Men'shchikov\inst{1},      
      Ph.~Andr\'e\inst{1},
       N.~Peretto\inst{1},        
      L.~D.~Anderson\inst{4,5},
      D.~Arzoumanian\inst{1},
      L.~Deharveng\inst{4},	
      P.~Didelon\inst{1},
      J.~Di~Francesco\inst{6},
     M.~J.~Griffin\inst{7},
     J.~M.~Kirk\inst{7},	
      V.~K\"onyves\inst{1},
      P.~G.~Martin\inst{8},
      A.~Maury\inst{9},
      V.~Minier\inst{1}.
      S.~Molinari\inst{2},            
      M.~Pestalozzi\inst{2},
       S.~Pezzuto\inst{2},
      M.~Reid\inst{8},
   H.~Roussel\inst{10},
   M.~Sauvage\inst{1},
         F.~Schuller\inst{11},
    L.~Testi\inst{9},
       D.~Ward-Thompson\inst{7},
      G.~J.~White\inst{12,13},
     A.~Zavagno\inst{4}}         
\institute{Laboratoire AIM Paris-Saclay, CEA/IRFU - CNRS/INSU - Universit\'e Paris Diderot, Service d'Astrophysique, B\^at. 709, CEA-Saclay, F-91191 Gif-sur-Yvette Cedex, France
\email{quang.nguyen-luong@cea.fr} \label{cea}
      \date{Received 05~Aug~2011; accepted 16~Sep~2011}
\and
INAF-IFSI, Via del Fosso del Cavaliere 100, 00133 Roma, Italy \label{ifsi}
\and 
Laboratoire d'Astrophysique de Bordeaux, CNRS/INSU -- Universit\'e de Bordeaux,
BP 89, 33271, Floirac cedex, France \label{lab}
\and
Laboratoire d'Astrophysique de Marseille , CNRS/INSU -- Universit\'e de Provence, 13388, Marseille cedex 13, France \label{lam}
\and
Physics Department, West Virginia University, Morgantown, WV 26506, USA
\and
National Research Council of Canada, Herzberg Institute of Astrophysics,
University of Victoria, Department of Physics and Astronomy, Victoria, Canada \label{can}
\and
Cardiff University School of Physics and Astronomy, UK \label{cardiff}
\and
CITA \& Dep. of Astronomy and Astrophysics, University of Toronto, Toronto, Canada \label{cita}
\and
ESO, Karl Schwarzschild Str. 2, 85748, Garching, Germany \label{eso}
\and
Institut d'Astrophysique de Paris, UMR7095 CNRS, Universit\'e Pierre \& Marie Curie, 98 bis Boulevard Arago, 75014 Paris, France 
\and 
Max-Planck-Institut f\"ur Radioastronomie, Auf dem H\"ugel 69, D-53121 Bonn, Germany \label{mpifr}
\and
The Rutherford Appleton Laboratory, Chilton, Didcot, OX11 0NL, UK
\and
Department of Physics and Astronomy, The Open University, Milton Keynes, UK
}

\titlerunning{Massive star formation in G035.39--00.33, a mini-starburst event}

\authorrunning{Q.~Nguy$\tilde{\hat{\rm e}}$n~Lu{\hskip-0.65mm\small'{}\hskip-0.5mm}o{\hskip-0.65mm\small'{}\hskip-0.5mm}ng et al.}

 \abstract
      {The filament IRDC G035.39--00.33 in the W48 molecular complex is one of the darkest infrared clouds observed by \textit{Spitzer}. It has been observed by the PACS (70 and 160\,$\micron$) and SPIRE (250, 350, and 500\,$\micron$) cameras of the \textit{Herschel} Space Observatory as part of the W48 molecular cloud complex in the framework of the HOBYS key programme. The observations reveal a sample of 28 compact sources (deconvolved FWHM sizes $<$0.3~pc) complete down to $\sim$$5~\msun$ in G035.39--00.33 and its surroundings. Among them, 13 compact sources are massive dense cores with masses $>$$20~\msun$. The cloud characteristics we derive from the analysis of their spectral energy distributions are masses of $20-50~\msun$, sizes of 0.1--0.2~pc, and average densities of $2-20 \times 10^{5}~\cmc$, which make these massive dense cores excellent candidates to form intermediate- to high-mass stars. Most of the massive dense cores are located inside the G035.39--00.33 ridge and host IR-quiet high-mass protostars. The large number of protostars found in this filament suggests that we are witnessing a mini-burst of star formation with an efficiency of $\sim$$15\%$ and a rate density of $\sim$$40~\msun\,$yr$^{-1}\,$kpc$^{-2}$ within $\sim$8~pc$^2$, a large area covering the full ridge. 
        Part of the extended SiO emission observed towards G035.39--00.33 is not associated with obvious protostars and may originate from low-velocity shocks within converging flows, as advocated by previous studies.
}
\maketitle

\section{Introduction}
The $\textit{Herschel}$ observatory \citep{2010A&A...518L...1P}, operating at far-infrared to submilimeter wavelengths, is an excellent tool for studying the early phases of star formation and the connection of star precursors to the ambient cloud. To address fundamental questions of massive star formation, the guaranteed-time key programme ``$\textit{Herschel}$ imaging survey of OB Young Stellar objects'' (HOBYS\footnote{\it http://hobys-herschel.cea.fr}) has observed all of the regions forming high-mass stars within a distance of 3~kpc from the Sun, one of which is the W48 molecular complex. This survey will ultimately provide an unbiased view of the formation of OB-type stars and the influence of the ambient environment on that process (see \citealt{2010A&A...518L..77M, 2010A&A...518L..83S, 2010A&A...518L..84H, 2010A&A...518L..91D, hill11}). 
A picture is starting to emerge that high-mass stars are formed from more dynamical processes than low-mass stars. 
Massive dense cores (MDCs, $\sim$0.1~pc, $>$$10^5~\cmc$) 
either IR-bright or IR-quiet, where the mid-infrared flux threshold is used as a proxy for the presence/absence of a high-mass stellar embryo (see 
\citealt{motte07}), have short lifetimes. The high-mass class~0-like protostars forming within IR-quiet MDCs \citep{bontemps10} are observed to be fed on small scales by supersonic gas flows \citep{2011A&A...527A.135C}. These short timescales and fast gas flows are consistent with molecular clouds and dense filaments dynamically formed by colliding flows of H {\scriptsize I} gas (e.g. \citealt{schneider10,Nguyen-Luong:2011uq}).
\begin{figure}[!htbp]
\centering
$\begin{array}{c}
  \resizebox{\hsize}{!}{\includegraphics[angle=0,width=8.2cm]{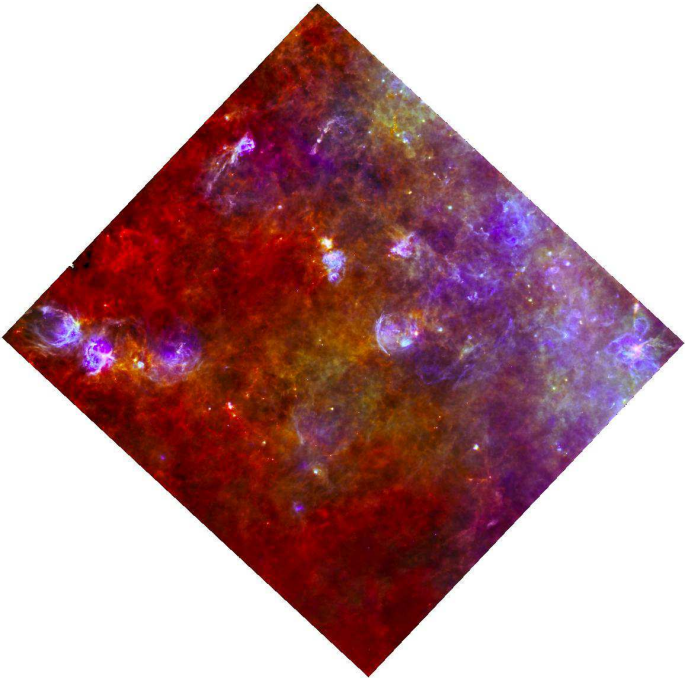}} \\
\end{array}$
\vskip -0.3cm
\caption{The three-colour image built from \textit{Herschel} images with red~= 250~$\micron$, green=160~$\micron$, and blue=70~$\micron$. The bright diffuse emission on the right corresponds to the Galactic plane. H~{\scriptsize II} regions are prominent blue loops/bubbles, while earlier stage star-forming sites are red filaments.
}
\label{fig:3colour}
\vskip -0.5cm
\end{figure}
Cold and dense filamentary structures in molecular clouds are potential sites to find the precursors of high-mass stars. One such example are Infrared dark clouds (IRDCs), which are dark extinction features against the Galactic background at mid-IR wavelengths. They have rather high column densities ($>$$10^{22}~ \cmd$), cold temperatures ($<$20~K), and filamentary structures (e.g. \citealt{2010ApJ...723..555P}),
bearing resemblances to the swept-up gas features in the simulations of colliding atomic gas (e.g. \citealt{heitsch09,banerjee09}).

The IRDC G035.39--00.33 (also called G035.49-00.30) is an IRDC filament located in the W48 molecular complex at a distance of $\sim$3~kpc \citep{2010A&A...515A..42R}. 
Using $^{13}$CO emission, \cite{2006ApJ...653.1325S} estimate that the filament has a mass of $\sim$9000 $\msun$ in an area with an effective radius of $\sim$10~pc and a median column density of $1 \times 10^{22}~$cm$^{-2}$. 
\cite{jimenez-serra10} observed that extended SiO emission is associated with this filament, which they interpreted as being produced by low-velocity shocks associated with colliding gas flows and/or shocks from protostellar outflows. Here we use $\textit{Herschel}$ data to investigate the star formation activity in G035.39--00.33 and determine whether any of the SiO emission could be associated with converging flows.

\section{$\textit{Herschel}$ observations and ancillary data}

The entire W48 molecular complex was observed on 18 and 19 September 2010 using PACS \citep{2010A&A...518L...2P} at $70/160\,\micron$ and SPIRE \citep{2010A&A...518L...3G} at $250/350/500\,\micron$ in the parallel scan-map mode with a scanning speed of 20\arcsec/s. Two perpendicular scans were taken to cover a SPIRE/PACS common area of $2.5\degr~ $x$~ 2.5\degr$. 
The data were reduced in two steps. The raw data (level-0) of the individual scans from both PACS and SPIRE were calibrated and deglitched using HIPE\footnote{HIPE is a joint development software by the Herschel Science Ground Segment Consortium, consisting of ESA, the NASA Herschel Science Center, and the HIFI, PACS, and SPIRE consortia.} pipeline version 7.0. The SPIRE and PACS level-1 data were then fed to version 4 of the Scanamorphos software package\footnote{\it http://www2.iap.fr/users/roussel/herschel/} (Roussel 2011, submitted), which subtracts brightness drifts by exploiting the redundancy of observed points on the sky, masks remaining glitches, and produces maps. 
The final images have angular resolutions of $\sim$6\arcsec, 12\arcsec, 18\arcsec, 25\arcsec, and 37\arcsec~and 1$\sigma$ rms of 0.02~Jy/1.4$\arcsec$-pixel, 0.08~Jy/2.8$\arcsec$-pixel, 1~Jy/beam, 1.1~Jy/beam, and 1.2~Jy/beam for 70\,$\micron$, 160\,$\micron$, 250\,$\micron$, 350\,$\micron$, and 500\,$\micron$, respectively. All \textit{Herschel} images were converted to MJy/sr by multiplying the aforementioned maps by 21706, 5237, 115, 60, and 27 for 70\,$\micron$, 160\,$\micron$, 250\,$\micron$, 350\,$\micron$, and 500\,$\micron$, respectively.   
The final maps of W48 are shown as a three-colour image in Fig.~\ref{fig:3colour} and individual images in Figs.~\ref{fig:imagesall}a-e.

\begin{figure*}[!bhtp]
$
\begin{array}{c}
 \includegraphics[angle=-0,width=19cm]{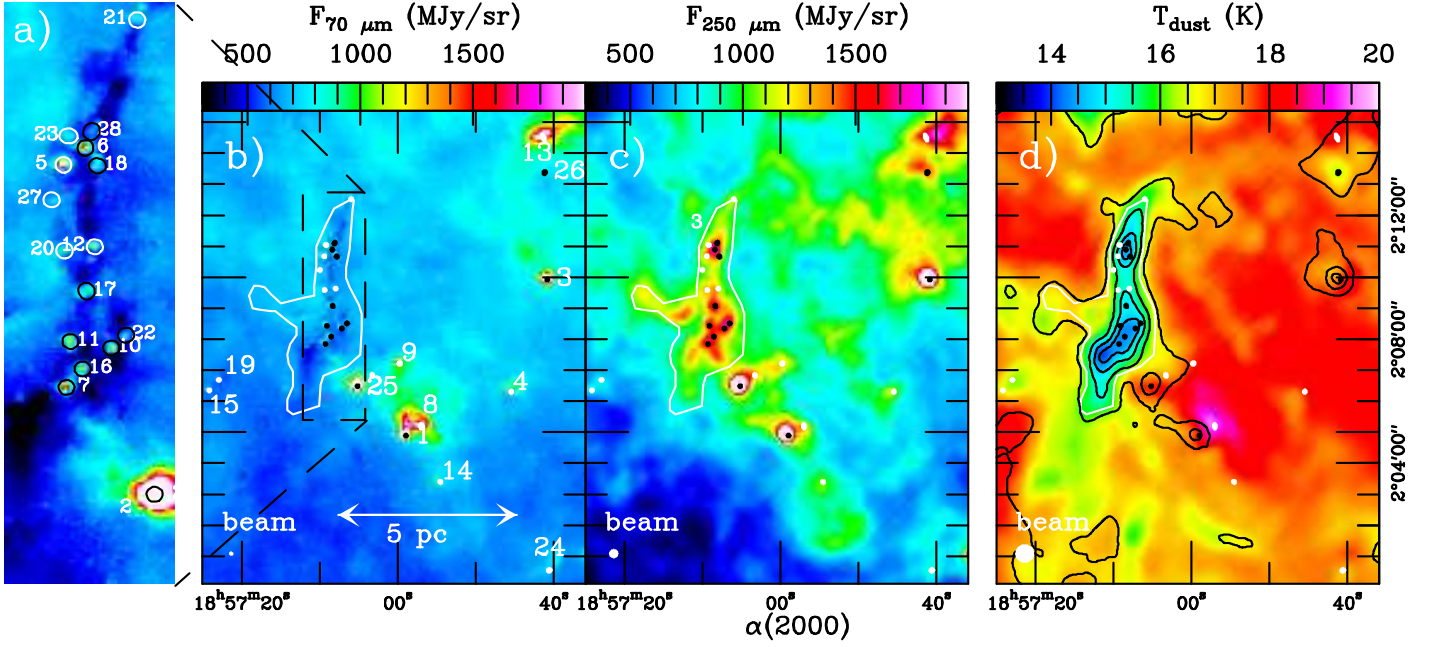}

 \end{array}
$
\vskip -1.2cm
\caption{{\bf (a)} PACS 70\,$\micron$ image of the G035.39--00.33 filament; ({\bf b)} PACS 70\,$\micron$ image of the G035.39--00.33 filament and its surroundings; ({\bf c)} SPIRE 250\,$\micron$ image of the G035.39--00.33 filament and its surroundings; ({\bf d)} Temperature (colour) and column density (contours from 1.5 to 9 by $1.5~\times~10^{22}~\cmd$) images. The dense cores  with mass $>$$20~\msun$ are indicated by black ellipses, those with mass $<$$20~\msun$ by white ellipses. 
The elliptical sizes represent the \textit{FWHM} sizes at 160~$\micron$.
The extent of the IRDC ($>$$3~\times~10^{22}~\cmd$) is indicated by a white polygon in {\bf b-d}.}

\label{fig:dens_T}
\end{figure*}
To complement the spectral energy distributions (SEDs) of young stellar objects, we have used $\textit{Spitzer}$ data at 3.6, 8, and 24~$\micron$ (from the GLIMPSE and MIPSGAL surveys; \citealt{benjamin03}, \citealt{2009PASP..121...76C})\footnote{The Galactic Legacy Infrared Mid-Plane Survey Extraordinaire (GLIMPSE) and the Multiband Imaging Photometer for \textit{Spitzer} GALactic plane survey (MIPSGAL) provide $3.6-24\,\mu$m images of the inner Galactic plane with $1.5\arcsec-18\arcsec$ resolutions. Detailed information and reduced images are available at {\it http://www.astro.wisc.edu/sirtf/} and {\it http://mipsgal.ipac.caltech.edu/}}, 
LABOCA data at 870~$\micron$ (from the ATLASGAL survey; \citealt{schuller09})\footnote{ The APEX Telescope Large Area Survey of the GALaxy (ATLASGAL) provides $870\,\mu$m images of the inner Galactic plane with $19\arcsec$ resolution. Reduced images will soon be available at \it http://www.mpifr-bonn.mpg.de/div/atlasgal/index.html},
 and BOLOCAM data at 1.1~mm (from the BGPS survey; \citealt{2010ApJ...721..137B})\footnote{ The BOLOCAM Galactic Plane Survey (BGPS) provides 1.1~mm images of the inner Galactic plane with $33\arcsec$ resolution. Reduced images are available at  \it http://irsa.ipac.caltech.edu/data/BOLOCAM\_GPS/}. 
We note that we used the images but not the compact source catalogues produced by these surveys since we extracted sources using a new algorithm simultaneously on all ten images (see Sect.~4.1).  

In the present study, we did not apply colour corrections for PACS, SPIRE, or ancillary data. These flux corrections would be rather small ($<$10\% in the case of \textit{Herschel}) and are covered by the 30\% absolute calibration uncertainty we use when fitting SEDs (see Sect.~4.3). 

\section{The G035.39--00.33 ridge characterized by $\textit{Herschel}$}
\label{irdcstructure}
\begin{figure*}[hbtp]
\centering
$\begin{array}{cc}
\includegraphics[angle=0,width=9.cm]{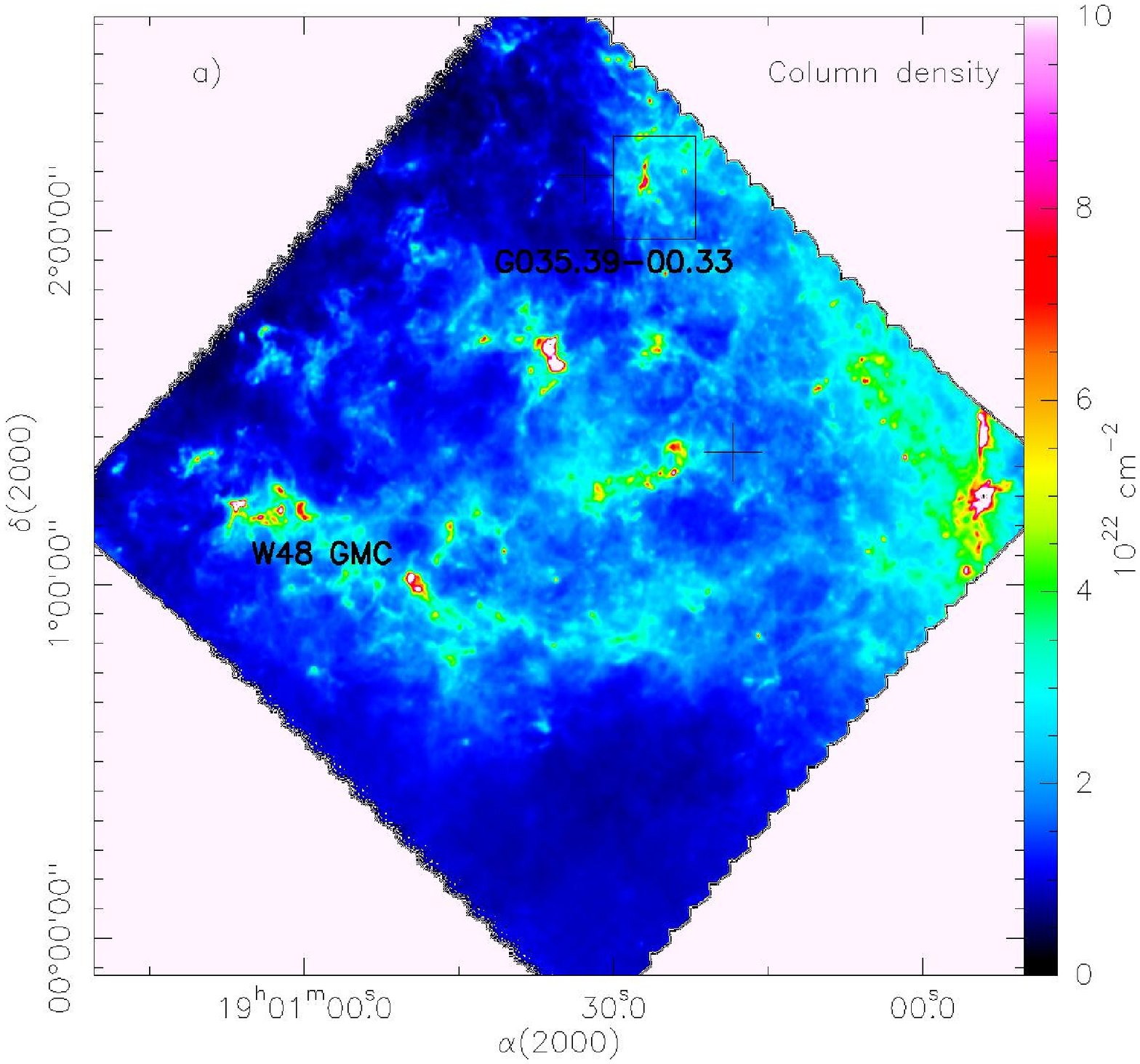}   & 
 \includegraphics[angle=0,width=8.4cm]{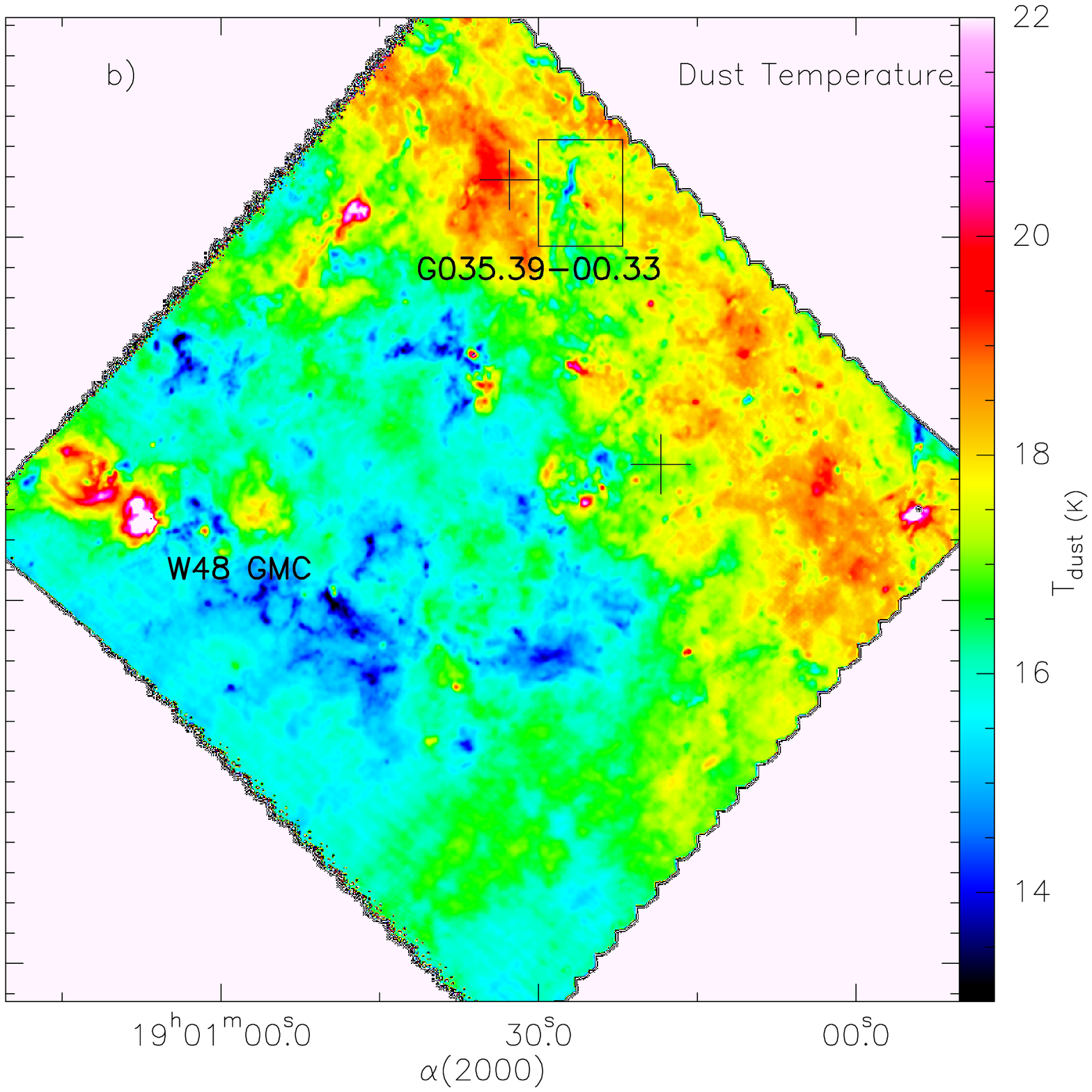} \\
\end{array}$
\vskip -0.3cm
\caption{Column density {\bf(a)} and dust temperature {\bf(b)} maps of W48 compiled from \textit{Herschel} images with a $HPBW$ of $37\arcsec$.  The dust opacity law of \cite{1983QJRAS..24..267H} is used with a dust emissitivity $\beta=2$. The W48 giant molecular cloud is indicated, the zoom on the IRDC G035.39--00.33 and its surroundings shown in Figs.~\ref{fig:dens_T}b-d is outlined. Plus signs (+) indicate the location of supernova remnants.}
\label{fig:Tdensall}
\end{figure*}

\begin{table*}[hbpt]
\vskip 0.5cm
\caption{Characteristics of the compact sources in G035.39--00.33 and its close surroundings: sources outside the IRDC are marked with an asterisk; \textit{FWHM} deconvolved sizes are measured at $160\,\micron$; dust temperature and mass are derived from SED modelling (see Fig.~\ref{fig:SEDex}); bolometric and submillimeter luminosities are measured from SEDs (see Sect.~\ref{sect:phy}) and lower limit of $\lsubmm/\lbol$ are given when sources are undetected at any wavelengths in the 3.6-70 micron range.}
\label{table:MDC} \centering \begin{tabular}{lllccrrcl } \hline
\hline
\# 	& ~~~RA~(2000)  &   Dec~(2000)   & Size	&     \tdust   &   $M$ & \lbol    &  $\lsubmm/\lbol$ & Possible nature\\   
	&   &     & [pc]	&	[K]	& [$\msun$]       &[$\lsun$]   &    [\%]  &  \\   
\hline
   {\bf 1*} 	& {\bf 18 : 56 : 59.0}  &   {\bf 2 : 04 : 53}   & {\bf 0.12} 	& {\bf 26$\pm$6} & {\bf 21$\pm$14} &   {\bf 3300} & {\bf 0.6} &  {\bf UCH~{\scriptsize II} region} \\
   {\bf 2*} 	& {\bf 18 : 57 : 05.1}  &   {\bf 2 : 06 : 29 }  & {\bf 0.12 }	&  {\bf 27$\pm$6 }& {\bf24$\pm$16 }& {\bf4700 }&  {\bf0.5 }& {\bf IR-bright protostellar MDC}\\
   {\bf 3*} 	& {\bf 18 : 56 : 40.9}  &   {\bf 2 : 09 : 55  } & {\bf 0.13 }	& {\bf 24$\pm$5 }& {\bf34$\pm$22 }&  {\bf 3100 }&{\bf  0.7 }& {\bf IR-bright protostellar MDC}\\ 
   4*  		& 18 : 56 : 45.5  &   2 : 06 : 18   &  0.13	&  25$\pm$9	 & 2$\pm$2   &   130   &  2     & low-mass dense core \\
   5		& 18 : 57 : 09.4  &   2 : 10 : 41   &  0.13	& 17$\pm$2 & 9$\pm$1   &    52--100 & $>$5 & low-mass dense core \\	    
   {\bf 6} 	& {\bf 18 : 57 : 08.4}  &  {\bf  2 : 10 : 53  } & {\bf 0.13 }	& {\bf16$\pm$3 }& {\bf20$\pm$12 }&  {\bf  70--200 }&{\bf $>$5 }&{\bf IR-quiet protostellar MDC}\\ 
   {\bf 7} 	& {\bf 18 : 57 : 09.3}  &  {\bf  2 : 07 : 51  } & {\bf 0.12 }	& {\bf12$\pm$1 }& {\bf49$\pm$17 }&  {\bf  50--130 }&{\bf $>$6 }&{\bf IR-quiet protostellar MDC}\\
   8*		& 18 : 56 : 57.0  &   2 : 05 : 11   &  0.20	& 23$\pm$8 & 5$\pm$4   &    130--210 & $>$1 & low-mass dense core \\	    
   9*		& 18 : 56 : 59.7  &   2 : 07 : 13   &  0.17	& 22$\pm$4 & 3$\pm$1   &    70--120 & $>$3 & low-mass dense core \\	    
  {\bf 10} 	& {\bf 18 : 57 : 07.2 } &  {\bf  2 : 08 : 21 }  & {\bf 0.12 }	& {\bf13$\pm$2 }& {\bf32$\pm$14 }&   {\bf 40--110 }&{\bf $>$8} &{\bf IR-quiet protostellar MDC}\\
  {\bf 11} 	& {\bf 18 : 57 : 09.1}  &  {\bf  2 : 08 : 26  } & {\bf 0.12 }	& {\bf 12$\pm$1 }&{\bf 37$\pm$20 }&   {\bf 40--100 }&{\bf $>$7 }&{\bf IR-quiet protostellar MDC}\\
  12		& 18 : 57 : 07.9  &   2 : 09 : 38   &  0.12	& 14$\pm$3 & 12$\pm$7 &    20--50 & $>$11 & low-mass dense core \\	    
  13*		& 18 : 56 : 41.4  &   2 : 14 : 30   &  0.27	& 21$\pm$8 & 7$\pm$5   &   210--280 & $>$2 & low-mass dense core \\	    
  14*		& 18 : 56 : 54.5  &   2 : 03 : 24   &  0.14	& 18$\pm$5 & 3$\pm$2   &    30--60 & $>$7 & low-mass dense core \\	    
  15*		& 18 : 57 : 24.1  &   2 : 06 : 21   &  0.13	& 21$\pm$2 & 2$\pm$1   &    20--40 & $>$4 & low-mass dense core \\	    
  {\bf 16} 	& {\bf 18 : 57 : 08.5}  &  {\bf  2 : 08 : 05  } & {\bf 0.12 }	& {\bf11$\pm$1} &{\bf 46$\pm$19 }& {\bf   30--90 }&{\bf $>$9 }&{\bf IR-quiet protostellar MDC}\\
  {\bf 17} 	& {\bf 18 : 57 : 08.3}  &  {\bf  2 : 09 : 04 }  & {\bf 0.15 }	& {\bf13$\pm$2 }&{\bf 50$\pm$19 }&  {\bf  50--140} &{\bf $>$9 }&{\bf IR-quiet protostellar MDC}\\
  {\bf 18} 	& {\bf 18 : 57 : 07.8}  &  {\bf  2 : 10 : 40 }  & {\bf 0.14 }	& {\bf14$\pm$2 }&{\bf 20$\pm$9 }&  {\bf  40--120} &{\bf $>$6 } &{\bf IR-quiet protostellar MDC}\\
   19*		& 18 : 57 : 23.0  &   2 : 06 : 41   &  0.12	& 15$\pm$2 & 3$\pm$1   &    10--20 & $>$9 & low-mass dense core \\	    
   20		& 18 : 57 : 09.4  &   2 : 09 : 35   &  0.13	& 14$\pm$2 & 16$\pm$5   &  30--60 & $>$10 & low-mass dense core \\	    
   21		& 18 : 57 : 06.0  &   2 : 12 : 31   &  0.13	& 16$\pm$2 & 5$\pm$3  &    20--30 & $>$7 & low-mass dense core \\	    
   {\bf 22}	& {\bf 18 : 57 : 06.5}  &  {\bf  2 : 08 : 31 }  &{\bf 0.13 }	& {\bf14$\pm$2} &{\bf 23$\pm$13 }&  {\bf 30--90} &{\bf $>$10 }&{\bf IR-quiet protostellar MDC}\\
   23		& 18 : 57 : 09.2  &   2 : 11 : 03   &  0.15	& 14$\pm$6 & 5$\pm$4  &    10--20 & $>$6 & low-mass dense core \\	    
   24*		& 18 : 56 : 40.6  &   2 : 00 : 33   &  0.18	& 15$\pm$5 & 5$\pm$4  &   20--30 & $>$6 & low-mass dense core \\	    
  25*		& 18 : 57 : 03.3  &   2 : 06 : 50   &  0.17	& 14$\pm$2 & 14$\pm$7  &  20--40 & $>$12 & low-mass dense core \\	       
  {\bf 26*}  & {\bf 18 : 56 : 41.0}  &  {\bf  2 : 13 : 22 }  & {\bf 0.19} & {\bf13$\pm$1} &{\bf 20$\pm$8 }& {\bf 20--40 }&{\bf $>$12} &{\bf Starless MDC}\\ 
  27		& 18 : 57 : 09.8  &   2 : 10 : 13   &  0.12	& - & -  &    5--10 & $>$35 & - \\	         
  {\bf 28}    & {\bf 18 : 57 : 08.1}  &  {\bf  2 : 11 : 06}   & {\bf 0.16} & {\bf11$\pm$1} &{\bf 55$\pm$11 }&  {\bf  30--60 }&{\bf $>$20} &{\bf IR-quiet protostellar MDC}\\
\hline
\hline
\end{tabular}
\begin{flushleft}
\end{flushleft}
\vskip -0.7cm 
\end{table*}

W48 is a massive molecular cloud complex ($8 \times 10^5~\msun$ that has been identified in the extinction map of Bontemps et al. in prep.) located slightly south of the Galactic plane. The numerous \hii regions (see Fig.~\ref{fig:3colour}) indicate that this region has actively formed high-mass stars. Figure~\ref{fig:3colour} also clearly displays future (high-mass) star-forming sites, such as the cold and dense filament IRDC G035.39--00.33,
 which is a prominent elongated structure oriented north-south in the $\textit{Herschel}$ maps (see Figs.~\ref{fig:dens_T}a-d). This site appears as a dark feature from $\textit{Herschel}$ 70~$\micron$ down to $\textit{Spitzer}$ 3.6~$\micron$, and in emission from 160~$\micron$ upward.

Prior to deriving any physical parameters, we added zero offsets, which were determined from the correlation with \textit{Planck} and \textit{IRAS} data following the procedure of \cite{2010A&A...518L..88B}. 
We then derived the dust temperature and column density maps of W48 (Figs.~\ref{fig:Tdensall}a-b) by fitting pixel-by-pixel single grey-body SEDs. Only the four longer-wavelength \textit{Herschel} bands, with a resolution of $37\arcsec$ and equal weight, were used in our SED fitting, as the 70~$\micron$ data may not 
be tracing the cold dust that we are most interested in. We assumed that the emission is optically thin at all wavelengths and took a dust opacity law of \cite{1983QJRAS..24..267H} with a dust emissivity index $\beta=2$ (see equation given in Sect.~\ref{sect:phy}).

The column density and temperature maps of Fig.~\ref{fig:dens_T}d show that the IRDC harbours dense and cold material ($\nhtwo\sim3-9\times10^{22}\,\cmd$ and $\tdust\sim13-16\,\mbox{K}$).
With a length of $\sim$6~pc and a 
width of $\sim$1.7~pc, this filamentary morphology is in good agreement with that found in the surface density map derived from 8~$\micron$ extinction images (e.g. \citealt{2009ApJ...696..484B}). 
The column density of G035.39--00.33 derived from \textit{Herschel} images is among the highest of known IRDCs \citep{2010ApJ...723..555P} and is typical of ``ridges", i.e. dominant filaments with a high column density as defined by \cite{hill11}. In contrast to most massive star-forming filaments, this IRDC has however a low temperature.
From the column density map, we estimate IRDC G035.39--00.33 to have a total mass of $\sim$5000\,$\msun$ within an area of 
 $\sim$8~pc$^{2}$ corresponding to A$_{V}>30$~mag as outlined in Fig.~\ref{fig:dens_T}d. This mass is similar to that derived from mid-infrared extinction \citep{2009ApJ...696..484B} and agrees well with that measured from a column density map of the IRDC where its background has been subtracted (method by \citealt{2010A&A...518L..98P}). In the following, we show that such a cold ridge has a high potential to form (massive) stars. 

\section{Massive dense cores in G035.39--00.33 and its surroundings} 
\subsection{Extracting \textit{Herschel} compact sources}

\textit{Herschel} compact sources were extracted using the multi-scale, 
multi-wavelength \textsl{getsources} algorithm (version 1.110614, see 
\citealt{2010A&A...518L.103M} and Men'shchikov in prep. for full 
details).
In our extraction, we first used \textit{only} the five 
\textit{Herschel} wavelength images for detecting sources and then 
\textit{all} ten wavelengths, from 3.6~$\micron$ to 1.1~mm, when measuring 
their fluxes.
At the detection step, the five \textit{Herschel} images were decomposed 
into multi-resolution cubes of single-scale single-wavelength images, 
as in the MRE--GCL method (see \citealt{motte07}). The five cubes were then combined, with greater weight being given to
the higher resolution images, into a single cube of single-scale 
combined-wavelength images. The \textit{Herschel} compact sources were 
finally detected within this cube, and their spatial positions and initial 
sizes were recorded.
At the measurement step and in the initial ten wavelength images 
successively, the sources properties (including \textit{FWHM} size 
and integrated flux) were computed at their detected location, after 
the background had been subtracted and overlapping sources had been 
deblended.
In the final \textsl{getsources} catalogue, each \textit{Herschel} compact 
source has a single position and ten \textit{FWHM} sizes at ten different wavelengths.

We selected sources with deconvolved \textit{FWHM} sizes at 160~$\micron$ smaller than 0.3~pc and a $>$$7\sigma$ detection at more than two \textit{Herschel} wavelengths (equivalent to $>$$5\sigma$ measurements at more than six wavelengths),
giving a catalogue of 28 sources complete down to 5~$\msun$ (see~Figs.~\ref{fig:dens_T}b-c and Table~\ref{table:MDC}). 
 Among them, 13 sources were identified as probably the best high-mass star precursor candidates ($M~>~20~\msun$), which we refer to as massive dense cores. 
 We consider these 28 sources as robust sources because 92\% of them were also identified by the {\it cutex} and {\it csar} algorithms (\citealt{2011A&A...530A.133M}; Kirk in prep.).

The sources studied here and more generally by the HOBYS key programme are cloud fragments that are strongly centrally concentrated but do not have firm outer boundaries when studied from subparcsec to parcsec scales. Their inner part can be fitted by a Gaussian with a $\sim$0.1~pc \textit{FWHM} size (e.g. \citealt{motte07}), which can only be resolved by the PACS cameras at the $0.7-3$~kpc distances of HOBYS molecular cloud complexes. From submillimeter observations, these MDCs have been found to be quasi-spherical sources with a radial density distribution very close to the $\rho(r)\propto r^{-2}$ law for $r\sim0.1-1$~pc (e.g. \citealt{2002ApJ...566..945B, 2002ApJS..143..469M}). This density power law and thus a mass radial power-law of $M(<r)\propto r$ are theoretically expected for dense cores formed by their own gravity, from singular isothermal spheres \citep{1977ApJ...214..488S} to protocluster clumps \citep{2000ApJ...542..964A}. For optically thin dust emission and a weak temperature gradient ($T(r)\propto r^{-q}$ with $q=0$ for starless or IR-quiet MDCs and $q=0.4$ for IR-bright MDCs), one expects the fluxes integrated within the apertures to vary (close to) linearly with the angular radius: $F(<\theta)\propto\theta$ for IR-quiet protostellar dense cores and $F(<\theta)\propto\theta^{0.6}$ for IR-bright protostellar dense cores (see e.g. \citealt{2001A&A...365..440M} for more details in the case of protostellar envelopes). We thus expect that the deconvolved sizes and integrated fluxes of the dense cores extracted here vary with the beam size and in practice increase from 160 to $500\,\mu$m, in line  
with the increase by a factor of $>$3 in the \textit{Herschel} \textit{HPBW}. This behaviour is indeed observed here (see Table 2), which we stress is independent of the compact source extraction technique used because it has also been observed for sources extracted by the MRE--GCL (\citealt{2010A&A...518L..77M}) and the Cutex (Giannini et al. submitted) algorithms. 
\begin{table}[!htbp] 
\caption{Scaled fluxes and quality of SED fits for the catalogue of Table.~\ref{table:MDC}}

\label{table:scaled} \centering \begin{tabular}{lccccl } \hline 
\hline 
Wavelength [$\micron$]       & 160  & 250   &   350     & 500\\   
 $HPBW$~[\arcsec]  & 12 & 18 & 25 & 37 \\
\hline 
 $<FWHM_{\mbox{\tiny{$\lambda$}}}^{\mbox{\tiny dec}}>$~[pc]$^{a}$                             & 0.18 & 0.30  &  0.40    &  0.60 \\ 
$\frac{F_{\mbox{\tiny scaled}}}{F_{\mbox{\tiny original}}}^{b}$    & -      &  53\% &  39\%   &  27\% \\
$\frac{\left|F_{\mbox{\tiny original}}-F_{\mbox{\tiny original,GB}}\right|}{F_{\mbox{\tiny original,GB}}}^{c}$        & -      &  90\% &  220\%   &  510\% \\
$\frac{\left|F_{\mbox{\tiny scaled}}-F_{\mbox{\tiny scaled,GB}}\right|}{F_{\mbox{\tiny scaled,GB}}}^{d}$        & -      &  15\% &  20\%   &  60\% \\
\hline 
\hline 
\end{tabular} 
\flushleft
Notes:\\
(a): The deconvolved FWHM sizes at wavelength $\lambda$ were calculated from the average FWHM sizes measured by \textsl{getsources} and the beam sizes as $FWHM_{\mbox{\tiny{$\lambda$}}}^{\mbox{\tiny dec}}=\sqrt{< FWHM_{\mbox{\tiny{$\lambda$}}} >^2-HPBW^2}  $

(b):  $F_{\mbox{\tiny original}}$ and $F_{\mbox{\tiny scaled}}$ are the original and scaled fluxes.

(c): $F_{\mbox{\tiny original,GB}}$ is the grey-body flux fitted to the original fluxes.

(d): $F_{\mbox{\tiny scaled,GB}}$ is the grey-body flux fitted to the scaled fluxes. 
\vskip -0.5cm 
\end{table}
\subsection{Scaling \textit{Herschel} fluxes for compiling SEDs} 
\label{app:scaling}

To limit the influence of different \textit{Herschel} resolutions and restrict ourselves entirely to the MDCs/dense cores size ($\sim$0.1~pc), we followed the procedure introduced by \cite{2010A&A...518L..77M}. We kept the PACS 70 and 160\,$\micron$ fluxes unchanged and linearly scaled the SPIRE 250, 350, and 500\,$\micron$ fluxes to the source size measured at 160\,$\micron$. This process assumes that (1) the size measured at 160\,$\micron$ reflects the spatial scale of the dense cores; (2) the 250, 350, and 500\,$\micron$ emission are mainly optically thin; and (3) the $F(\textless\theta)\propto\theta$ relation mentioned in Sect.~4.1 applies. In contrast to the $\ge$160\,$\micron$ emission, compact 70\,$\micron$ emission originates from hot dust close to the protostar and does not trace the cold component of a dense core. 
The relation $F(\textless\theta)\propto\theta$ is correct if the density and temperature power laws mentioned above apply to a large portion of the dense cores, roughly from half the resolution at 160\,$\micron$ to the resolution at 500\,$\micron$, i.e. from 0.02-0.09~pc to 0.1-0.6~pc depending on the cloud distance. However it does not preclude the subfragmentation, inner density flattening, and/or inner heating of the (massive) dense cores. As explained in Sect.~4.1, these are reasonable assumptions for MDCs. In contrast, if an H\,{\small II} region develops, the density and temperature structures of its parent dense core will probably be strongly modified and such a scaling cannot apply. 
The \textit{Herschel} fluxes (and other submillimeter measurements here) are scaled following the simple equation
\begin{equation}
F^{\mbox{\tiny SED}}_{\mbox{\tiny{$\lambda$}}} = F_{\mbox{\tiny{$\lambda$}}} \times \frac{ FWHM_{160}^{\mbox{\tiny dec}} } {FWHM_{\mbox{\tiny{$\lambda$}}}^{\mbox{\tiny dec}}},
\end{equation}
 where $F_{\mbox{\tiny{$\lambda$}}}$ and $FWHM_{\mbox{\tiny{$\lambda$}}}^{\mbox{\tiny dec}}$ are the measured flux and deconvolved $FWHM$ size at wavelengths $\lambda\ge 250\,\micron$. For simplicity and because their characteristics do not impact our discussion, we kept a linear scaling for IR-bright MDCs \#1, \#2, and \#3.

\begin{figure}[!h]
\centering
$
\begin{array}{ccc}
\resizebox{\hsize}{!}{\includegraphics{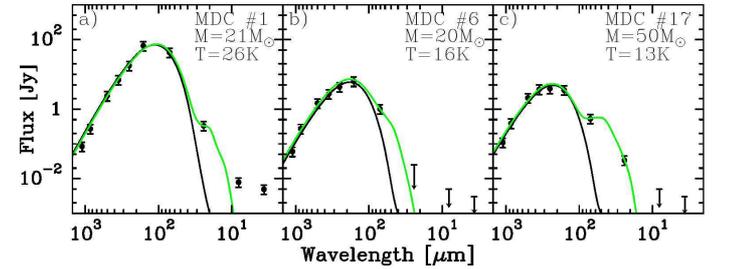}} 
\end{array}
$
\vskip -0.3cm
\caption{
SEDs built from \textit{Herschel} and other wavelengths of: the UCH~{\scriptsize II} region {\bf (a)} and two MDCs associated with SiO emission {\bf (b \& c)}. The curves are grey-body models fitted to data at wavelengths $\ge$~160~$\micron$. The single temperature grey-body fit (black curve) is consistent with the two temperatures grey-body fit (green curve). Error bars correspond to 30\% of the integrated fluxes.
}
\label{fig:SEDex}
\vskip 0.3cm
\end{figure}

\begin{figure}[htbp]
\centering
\includegraphics[angle=0,width=5.cm]{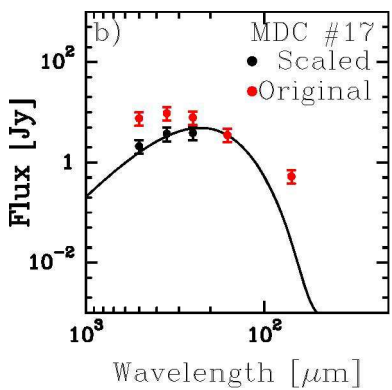} 
\vskip -0.1cm
\caption{Example SED compiled from fluxes with (in black) and without (in red) scaling for MDC \#17. The SED with scaled fluxes portrays a more classical behaviour for the Rayleigh-Jeans part of the protostellar dense core.}
\label{fig:SEDscaled}
\end{figure}

\begin{figure}[!hbpt]
\resizebox{\hsize}{!}{\includegraphics{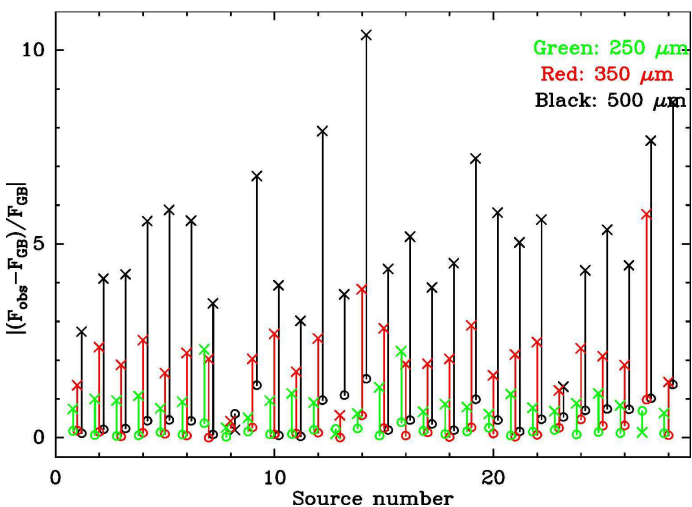}}
\vskip -0.1cm
\caption{The absolute value of the normalized difference between the original fluxes with their best-fit grey-body fluxes (crosses) and between the scaled fluxes with their best-fit grey-body fluxes (circles).}
\label{fig:fluxdifference}
\end{figure}

The complete SEDs of the 28 dense cores of Table~\ref{table:MDC} are built from \textit{Herschel} and ancillary data (see e.g. Fig.~\ref{fig:SEDex}). They can be compared to grey-body models if they arise from a unique gas mass. The scaling procedure used here allows for this, for a mass reservoir defined at $160\,\mu$m. Grey-body models are not perfect models but closely represent the long wavelength component ($160\,\micron$ to 1.1~mm) of SEDs compiled for sources such as starless, IR-quiet, or IR-bright protostellar MDCs. For the example MDC \#17, without scaling, \textit{Herschel} provides almost constant fluxes from $250\,\micron$ to $500\,\micron$ (see Fig.~\ref{fig:SEDscaled}), making the grey-body fit very unreliable. A few other SEDs even increase from $160\,\micron$ to $500\,\micron$, unrealistically suggesting a $<$5--10~K dust temperature. Without complementary submillimeter observations of high angular resolution and without any scaling, the SED of several cloud fragments, such as MDC \#17, could not be fit at all. As shown in Fig.~\ref{fig:SEDscaled}, the scaled SPIRE fluxes portray a more classical behaviour for the Rayleigh-Jeans part of a protostellar dense core SED. 
As can be seen in Table~\ref{table:scaled}, the $250\,\micron$, $350\,\micron$, and $500\,\micron$ fluxes are quantitatively lower but these flux changes help to improve the SED fit by a grey-body model. Indeed, the variations between the measured \textit{Herschel} fluxes and the fitted grey-body fluxes are $\sim$$5-10$ times larger in the case of the original fluxes than for scaled fluxes (see Fig.~\ref{fig:fluxdifference}).

\begin{figure}[htbp]
\resizebox{\hsize}{!}{\includegraphics{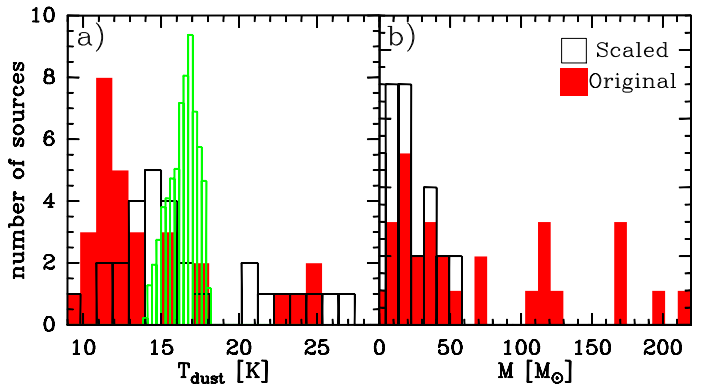}} 
\vskip -0.1cm
\caption{Dust temperature {\bf(a)} and mass {\bf(b)} distributions of the 28 compact sources derived from fitting a grey-body model to the original (in black) and scaled (in red) fluxes at wavelengths $\lambda\ge 160\,\micron$. The green histogram is the temperature distribution of the filament IRDC G0.35.39-00.33 taken from Fig.~\ref{fig:dens_T}d.}
\label{fig:ttextitisto}
\end{figure}

After flux scaling, the dust temperatures obtained from the grey-body models of sources within G035.39--00.33 are distributed around $<$\tdust$>\sim 14.5~$K (see Fig.~\ref{fig:ttextitisto}a).  They are in closer agreement with those measured from the dust temperature image ($\sim$16.5~K, see Fig.~\ref{fig:dens_T}d and \ref{fig:ttextitisto}a) and are $\sim$20\% higher than those obtained from grey-body models for the original fluxes ($\sim$11~K).
Most dust temperatures derived from SED fits to the original SPIRE fluxes are too cold. We do expect the temperature of compact cloud fragments to differ from that of their surrounding cloud and in practice to be lower if the protostellar dense cores are barely evolved as in IR-quiet massive dense cores. However, a $\sim$5~K  drop in temperature is quite extreme when averaged over only a three times smaller cloud fragment and a $<$\tdust$>\sim 11.3~$K for MDCs is unusually cold. For instance, a temperature of $\sim$20~K has been measured in Cygnus~X massive dense cores from ammonia measurements and \textit{Herschel} SEDs (\citealt{motte07}; Hennemann et al. in prep.).

\subsection{Deriving physical parameters for the compact sources}
\label{sect:phy}
After scaling the fluxes, we compiled a SED and fit a single grey-body to the 160\,$\micron$-1~mm wavelength data (see Figs.~\ref{fig:SEDex}a-c and \ref{fig:SEDex2}a-e) to derive the mass and the average dust temperature of each dense core. We assumed a dust opacity law $\kappa_\lambda=\kappa_{0}\times\left( \frac{\lambda}{300~\mu\mbox{\tiny m}} \right)^{-\beta}$ with $\kappa_{0}=0.1~\gcm$ and a dust emissivity index of $\beta=2$. 
Our limited knowledge of the dust emissivity makes masses systematically uncertain by a factor of $\sim$2. The mass and dust temperature uncertainties quoted in Table~\ref{table:MDC} reflect those of SED fits to fluxes varying by 30\% with a constant emissivity. We successfully counterchecked our results obtained with single grey-body fits to the long wavelength ($>$$160\,\mu$m) component of SEDs by fitting two-temperature grey-body models to the data (see Fig.~\ref{fig:SEDex}).

We also derived the bolometric luminosity $\lbol$ and submillimeter luminosity $\lsubmm$ by integrating fluxes below the measurement points i.e., from 3.6\,$\micron$ to 1.1~mm and from 350\,$\micron$ to 1.1~mm, respectively. 
When sources were not detected at mid-infrared wavelengths, Table~\ref{table:MDC} gives a range of bolometric luminosities using the 3.6, 8, and 24~$\micron$ noise levels as upper limits or setting them to 0.

Cloud fragments in our sample have masses ranging from 1\,$\msun$ to $\sim$50\,$\msun$ with deconvolved \textit{FWHM} sizes at 160\,$\micron$ ranging from 0.1 to 0.3~pc and temperatures from 10 to 30~K. The mass of the gas reservoir selected at $160\,\micron$ and measured by a grey-body fit is, on average, 40\% that obtained from the SED fit of original fluxes (see Fig.~\ref{fig:ttextitisto}b). 
The shrinking, by a factor of $\sim$2, of the mass reservoir arises from the decision to focus on the gas contained within the $160\,\micron$ sizes, which are twice as small as the mean sizes at the SED peak (generally close to $250\,\micron$). Without such a scaling, the present study would have considered cloud fragments that are, on average, twice as massive but also $\sim$8 times less dense. 
These fragments would thus not correspond as well to the typical scale and density of MDCs and would logically have a less efficient transfer of matter from the clump gas to star.
For the reasons given in Sects.~4.1-4.2 and despite the modifications implied, we estimate that the rescaling of SPIRE fluxes is a necessary step in a simple SED fitting approach applied to large samples of compact sources.

\section{Discussion}
\subsection{A burst of star formation activity}
\label{evolutionarystage}
We have focused here on the 13 best high-mass star precursor candidates, or MDCs, which are the most massive ($M> 20~\msun$) and dense (\textit{FWHM} averaged density of $>$$2 \times 10^{5}~\cmc$) cloud fragments, and thus the most compact (\textit{FWHM}$\,<0.19$~pc). Several MDCs lie within the clumps identified by \cite{2006ApJ...641..389R} at 1.3~mm, which are larger (0.2-0.9~pc) cloud structures.

The MDC \#1 was classified as an Ultra-Compact H~{\scriptsize II} (UCH~{\scriptsize II}) region based on its free-free emission at 1.4 and 5~GHz \citep{1994ApJS...91..347B}. The MDC \#2 is an UCH~{\scriptsize II} candidate according to the Red MSX Source survey \citep{2011A&A...525A.149M} but no centimeter free-free emission has yet been detected. Water maser emission was detected towards MDC \#2 \citep {2009ApJS..181..360C}. These two MDCs together with MDC \#3 are the most luminous MDCs in our sample, i.e, $\lbol\sim3000-5000~\lsun$, while others have $\lbol<300~\lsun$. The high luminosity of MDCs \#1, \#2, and \#3, their warm dust and high mass imply that they are already evolved massive star-forming sites such as IR-bright MDCs (\#2 and \#3) or UCH~{\scriptsize II}s (\#1). These three MDCs also have distinct 24\,$\micron$ fluxes ($>$1~Jy) compared to the remaining cores ($<$1~Jy). This flux limit is close to that used to define IR-quiet MDCs (see \citealt{motte07} after distance correction). 
The separation based on 24\,$\micron$ fluxes is also consistent with the $\lsubmm / \lbol$ ratio, which is small ($<$~1\%) for IR-bright MDCs or UCH~{\scriptsize II} regions and large ($>$~1-20\%) for IR-quiet MDCs harbouring high-mass class~0s or for starless dense core candidates. 
 \cite{2000prpl.conf...59A} used this ratio to separate low-mass class~0 and class~I protostars (size~$\sim0.01-0.1$~pc) with 1\% as the dividing line. 
Since MDCs are larger-scale cloud structures ($\sim0.1$~pc), the $\lsubmm / \lbol$ ratio of IR-quiet MDCs is expected to be substantially larger than those of low-mass class~0s.
Among the remaining ten MDCs with low 24~$\micron$ emission, one (MDC \#26, outside the G035.39--00.33) is a good candidate starless dense core since it has no emission at wavelengths shorter than 70\,$\micron$.

Interestingly, 55\% (15) of the dense cores and 70\% (9) of the MDCs are located inside the filament G035.39--00.33, which has a mass density of $600~\msun$\ pc$^{-2}$ and a mass per unit length of $800~\msun$\ pc$^{-1}$. It is much higher than those of the highly extincted regions (A$_{V}>30$~mag) of Perseus and Ophiuchus (see \citealt{2010ApJ...723.1019H}).
This high concentration of (massive) dense cores indicates that the gravitational potential of G035.39--00.33 helped them to build up. From their SED, the MDCs in the G035.39--00.33 filament are IR-quiet protostellar dense cores and thus harbour young protostars (see Table~\ref{table:MDC}). This predominance of IR-quiet MDCs suggests that most of these dense cores have been formed simultaneously and are probably forming just following the formation of the filament. In agreement with the presence of an UC\hii region and previous studies (e.g. \citealt{motte07}; \citealt{2010A&A...515A..55R}), we assumed a $20-40\%$ mass transfer from MDC to stellar mass. We therefore estimate that between four and nine of the MDCs (those with $>$$20-40~\msun$, see Table~1) would indeed form high-mass stars in G035.39--00.33. With nine high-mass protostars forming in this $5000~\msun$ filament and assuming the initial mass function of \cite{kroupa01}, the star formation efficiency (SFE) could be as high as $\sim$15\%.
The numerous protostars (being detected at $70\,\micron$) observed within the G035.39--00.33 filament (area of $\sim$8~pc$^{2}$) shown in Fig.~\ref{fig:dens_T}a, are estimated to have a mean mass on the main sequence of $\sim$$2~\msun$ (meaning that stars form in $5-50~\msun$ dense cores with 25\% mass transfer efficiency) and a lifetime of $\sim$10$^5$~yr, close to the dense-core free-fall time.
This implies a star formation rate (SFR) of $\sim$300~\msun\ Myr$^{-1}$, which exceeds the mean SFR of Gould Belt clouds by a factor of seven (see \citealt{2010ApJ...723.1019H}) and is close to that of the W43 mini-starburst region \citep{motte03, Nguyen-Luong:2011uq}. The SFR density within this $\sim$8~pc$^{2}$ high column-density (A$_{V}>30$~mag) filament, $\Sigma_{SFR}\sim 40~\msun\,$yr$^{-1}\,$kpc$^{-2}$, is high and has only been measured in Gould Belt dense clumps (10--100 times smaller areas with A$_{v}$$>$20--40 mag, see \citealt{2010ApJ...723.1019H,2011arXiv1108.0668M}) and in extragalactic circumnuclear starburst regions (e.g. \citealt{1998ApJ...498..541K}).

The measurements (SFE, SFR, $\Sigma_{SFR}$) of this high star formation activity are uncertain by a factor $>$2, coming essentially from the uncertainty in the number of high-mass protostars and the mass transfer efficiency from dense cores to stars (see above). In contrast, the flux scaling should have a relatively low influence on our results. Indeed, without scaling, massive clumps are twice as massive but would form stars of the same stellar mass on the main sequence. Despite these limitations, the SFE, SFR, and $\Sigma_{SFR}$ values above suggest that a mini-starburst event, i.e. a miniature model of events in starburst galaxies, is occurring in this filament. To keep the SFE in the entire W48 complex consistent with those seen in other clouds (1-3\% per 10$^{7}$~yrs, \citealt{silk97}) and reconcile the SFE and SFR values given above, the star formation event should be short ($\sim$10$^{6}$~yr). 
If confirmed, this mini-burst of recent star formation would be consistent with the filament G035.39--00.33 being formed through a rapid process such as converging flows. Another interpretation could be that star formation has recently been triggered, but a nearby triggering source has not been found.

\begin{figure}[hbtp]
\begin{center}
\resizebox{\hsize}{!}{\hspace{1mm}}
\vskip -0.7cm
\centering
\includegraphics[angle=0,width=8.5cm,viewport=0 0 300 350]{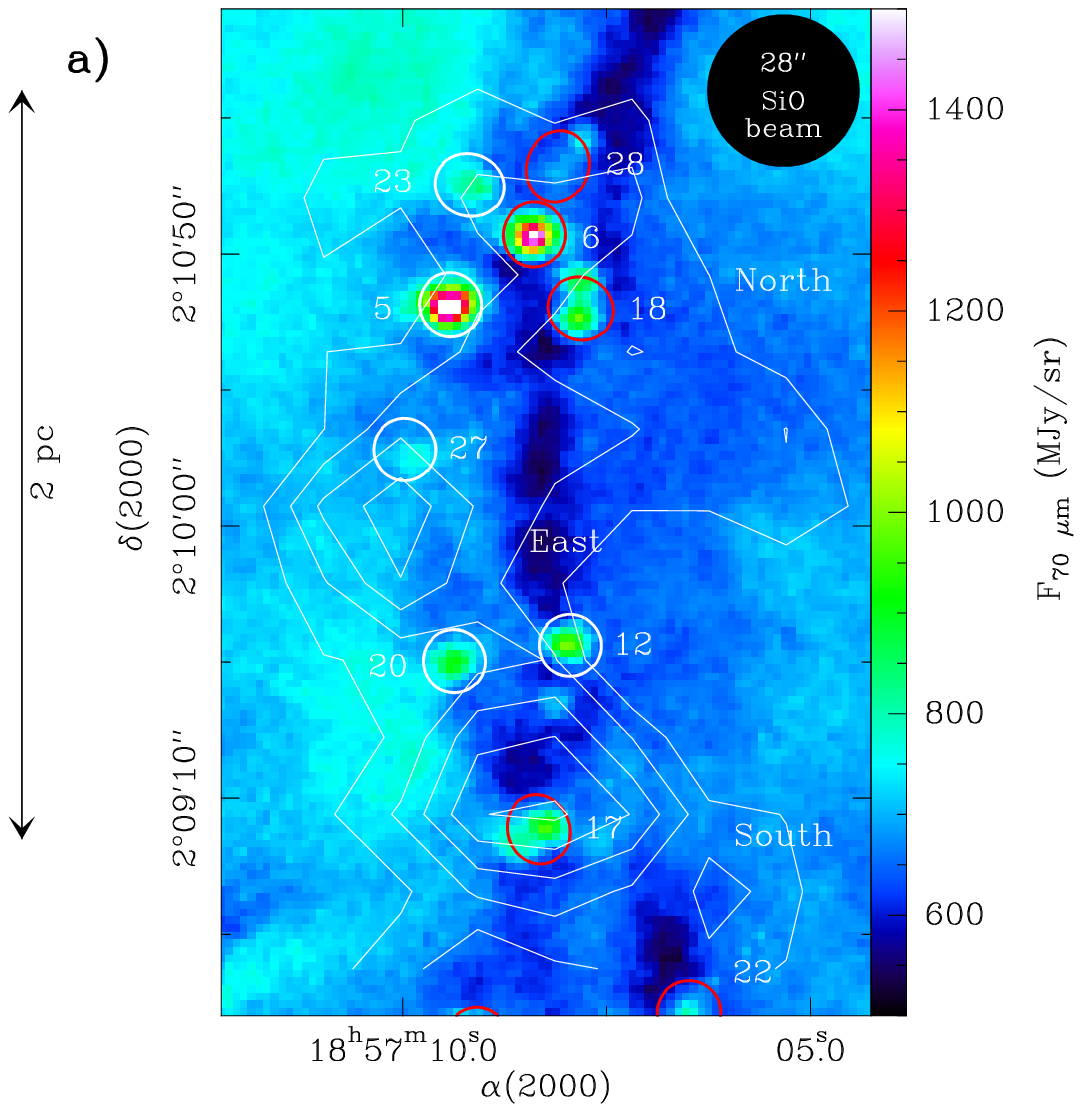} \\
\vskip -0.9cm
\hskip -5.5cm

\includegraphics[angle=0,width=11cm,viewport=0 0 420 100]{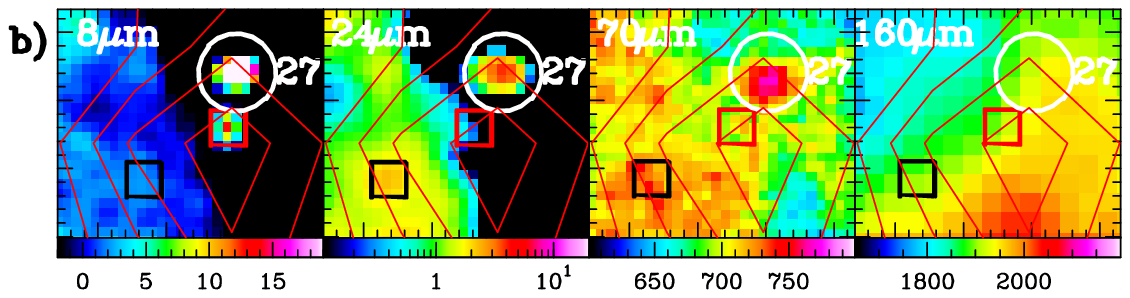} 
\vskip -0.1cm
\caption{ 
({\bf a}) Part of G035.39--00.33 seen at 70\,$\micron$ (colour) and in SiO (contours from 0.01 to 0.05~K$\cdot\kms$ by 0.01~K$\,\kms$ from \citealt{jimenez-serra10}). The dense cores  with mass $>$$20~\msun$ are indicated by red ellipses, those with mass $<$$20~\msun$ by white ellipses. ({\bf b}) Zoom towards source \#27 seen at 8--160~$\micron$. Red and black boxes indicate the 8~$\micron$--only and 24~$\micron$--only sources, respectively.}
\label{fig:Sio}
\end{center}
\vskip -0.5cm
\end{figure}
\subsection{Are protostars responsible for the SiO emission?}

\cite{jimenez-serra10} mapped the filament G035.39--00.33 in SiO~(2--1) (HPBW$\,\simeq28\arcsec$, see Fig.~\ref{fig:Sio}) finding $0.01-0.05$~K$\,\kms$ widespread emission along part of this filament. They proposed that it could be produced by low-velocity shocks associated with converging flows and/or outflows from protostars. SiO is indeed an excellent tracer of shocked gas associated with molecular outflows (e.g. \citealt{1997A&A...321..293S})
and is particularly prevalent towards early-stage high-mass protostars (e.g. \citealt{2001A&A...372..281H,motte07}). It has also been proposed
that low-velocity shocks in a high-density medium could sputter dust grains and enhance SiO in the gas phase (Lesaffre,~P., Gusdorf, A.~priv.~comm.).

We investigated the origin of this SiO emission by correlating its spatial distribution with that of young protostars detected by \textit{Herschel} (see Sect.~\ref{evolutionarystage} and Table~\ref{table:MDC}). Two of the three peaks of SiO (North and South) 
coincide with IR-quiet massive dense cores: MDCs \#6 and \#17
(see Figs.~\ref{fig:Sio} and \ref{fig:sionorthsouth}). Given that the SED of MDC \#17 displays a dominating cold and massive envelope and a warm component detected with \textit{Spitzer} (see Fig.~\ref{fig:SEDex}c), it most probably hosts an early-stage high-mass protostar. 
The MDC \#6 is not detected at wavelengths shorter than $24~\micron$ (see Fig.~\ref{fig:SEDex}b) but its compact $70~\mu$m emission is three times that of the flux predicted by the grey-body model that fits its cold envelope. This MDC also coincides with a 40$\arcsec$-beam water maser source \citep{2010A&A...515A..42R} suggesting that it harbours an intermediate- or high-mass protostar. The SiO emission at these two positions could easily originate from shocks within protostellar outflows because they are comparable to the weakest ones associated with IR-quiet MDCs in Cygnus~X (\citealt{motte07} after distance correction). 

In contrast, the eastern SiO peak, 
although detected in SiO~(2--1) at $\sim$0.04~K$\,\kms$, does not coincide with any reliable protostar candidate. Source \#27, which lies within the SiO beam, is only detected shortwards of 160\,$\micron$ (see Fig.~\ref{fig:Sio}b). A specific run of \textsl{getsources}, allowing low-reliability measurements (a $3\sigma$ detection at a single Herschel wavelength), has in fact been done at the location of the eastern SiO peak to extract source \#27. The SED of this source (see Fig.~\ref{fig:SEDex2}d) consists of two unrelated components: one mid-IR source of unknown nature detected at 8, 24, and 70~$\micron$ plus some nearby cloud structure whose 160~$\micron$-1.1~mm emission extends until the location of source \#27. Re-examining the \textit{Herschel} and \textit{Spitzer} images, we estimate that source \#27 could be at most a 1~$\msun$ dense core containing an evolved low-mass class I protostar. Since SiO outflows are known to be rare and confined to the proximity of intermediate-mass class 0 and high-mass protostars (e.g. \citealt{1997ApJ...482L..45M}; \citealt{1998A&A...333..287G}; \citealt{motte07}), there is only a low probability
that such a protostar 
could drive a strong-enough outflow to
be responsible for the 
SiO emission of the eastern peak. Two other sources observed within the SiO beam are only detected at 8\,$\micron$ or 24\,$\micron$ and are thus probably even more evolved young stellar objects or variations in the background emission. 
Outflows from neighbouring low- to high-mass protostars, $>$0.7~pc away from the eastern SiO peak, would also be unable to create such a SiO emission. Indeed, the extreme SiO outflow arising from the intermediate-mass star L1157 has emission peaks that are at less than 0.2~pc from the protostar \citep{1998A&A...333..287G}. The protostars here are unlikely to be extreme and less confined.

Since no definite protostar is detected by \textit{Herschel} towards the eastern SiO peak, the interpretation that the SiO emission could partly originate from large-scale converging-flows may be valid, at least at this location. Higher-resolution and higher-sensitivity observations would be necessary to definitively reach this conclusion. 
Further studies of the SiO emission and its relation to other high-density gas tracers, as well as the dust properties, will be presented by Henshaw et al. (in prep).

\vskip 0.3cm
\section{Conclusion}
We have used PACS and SPIRE maps of \textit{Herschel} to investigate the star formation activity in the IRDC G035.39--00.33, a cold ($13-16$~K) and dense ($\nhtwo$ up to $9\times 10^{22}~\cmd$) filament, which we have qualified as a ``ridge''. We have proposed a new approach to analysing the SED of compact sources compiled from \textit{Herschel} fluxes. We fitted single grey-body models to fluxes that had been linearly scaled downwards to correspond more closely to the mass reservoir emitting at $160\,\mu$m, the wavelength at which emission from the cold envelope of compact source is probed with highest angular resolution. On the basis of this procedure, we have found that \textit{Herschel} detected a total of 28 dense cores (\textit{FWHM}$\sim$0.15~pc), among them 13 MDCs with masses of 20--50~$\msun$ and densities of 2--20 $\times 10^{5}~\cmc$, which are potentially forming high-mass stars. Given their concentration in the IRDC G035.39--00.33, they may be participating in a mini-burst of star formation activity with 
 \textit{SFE}$\sim$ $15\%$, 
\textit{SFR}$\sim300~\msun\,$Myr$^{-1}$, and $\Sigma_{SFR}\sim 40~\msun\,$yr$^{-1}\,$kpc$^{-2}$. 
Two IR-quiet MDCs could be the origin of most of the extended SiO emission observed, the remainder possibly originating from a low-velocity shock within converging flows.

\begin{acknowledgements}
SPIRE has been developed by a consortium of institutes led by
Cardiff Univ. (UK) and including Univ. Lethbridge (Canada);
NAOC (China); CEA, LAM (France); IFSI, Univ. Padua (Italy);
IAC (Spain); Stockholm Observatory (Sweden); Imperial College
London, RAL, UCL-MSSL, UKATC, Univ. Sussex (UK); Caltech, JPL,
NHSC, Univ. colourado (USA). This development has been supported
by national funding agencies: CSA (Canada); NAOC (China); CEA,
CNES, CNRS (France); ASI (Italy); MCINN (Spain); SNSB (Sweden);
STFC (UK); and NASA (USA). \\
PACS has been developed by a consortium of institutes led by MPE (Germany)
and including UVIE (Austria); KU Leuven, CSL, IMEC (Belgium); CEA, LAM
(France); MPIA (Germany); INAF-IFSI/OAA/OAP/OAT, LENS, SISSA (Italy);
IAC (Spain). This development has been supported by the funding agencies
BMVIT (Austria), ESA-PRODEX (Belgium), CEA/CNES (France), DLR (Germany),
ASI/INAF (Italy), and CICYT/MCYT (Spain). \\
Part of this work was supported by the ANR (\textit{Agence Nationale pour la Recherche})
project ``PROBeS", number ANR-08-BLAN-0241. 
Tracey Hill is supported by a CEA/Marie-Curie Eurotalents Fellowship.
Kazi Rygl is supported by an ASI fellowship under contract number I/005/07/1.
We thank Izaskun Jim\'enez-Serra for providing the SiO image.
We thank the anonymous referee, whose comments contributed to improve the manuscript.
\end{acknowledgements}
\bibliographystyle{aa}
\bibliography{w48}	
\newpage

\newpage
\section{Online figures} 
\newpage
\begin{figure*}[hbtp]
\centering
$\begin{array}{cc}
\includegraphics[angle=0,width=8.cm]{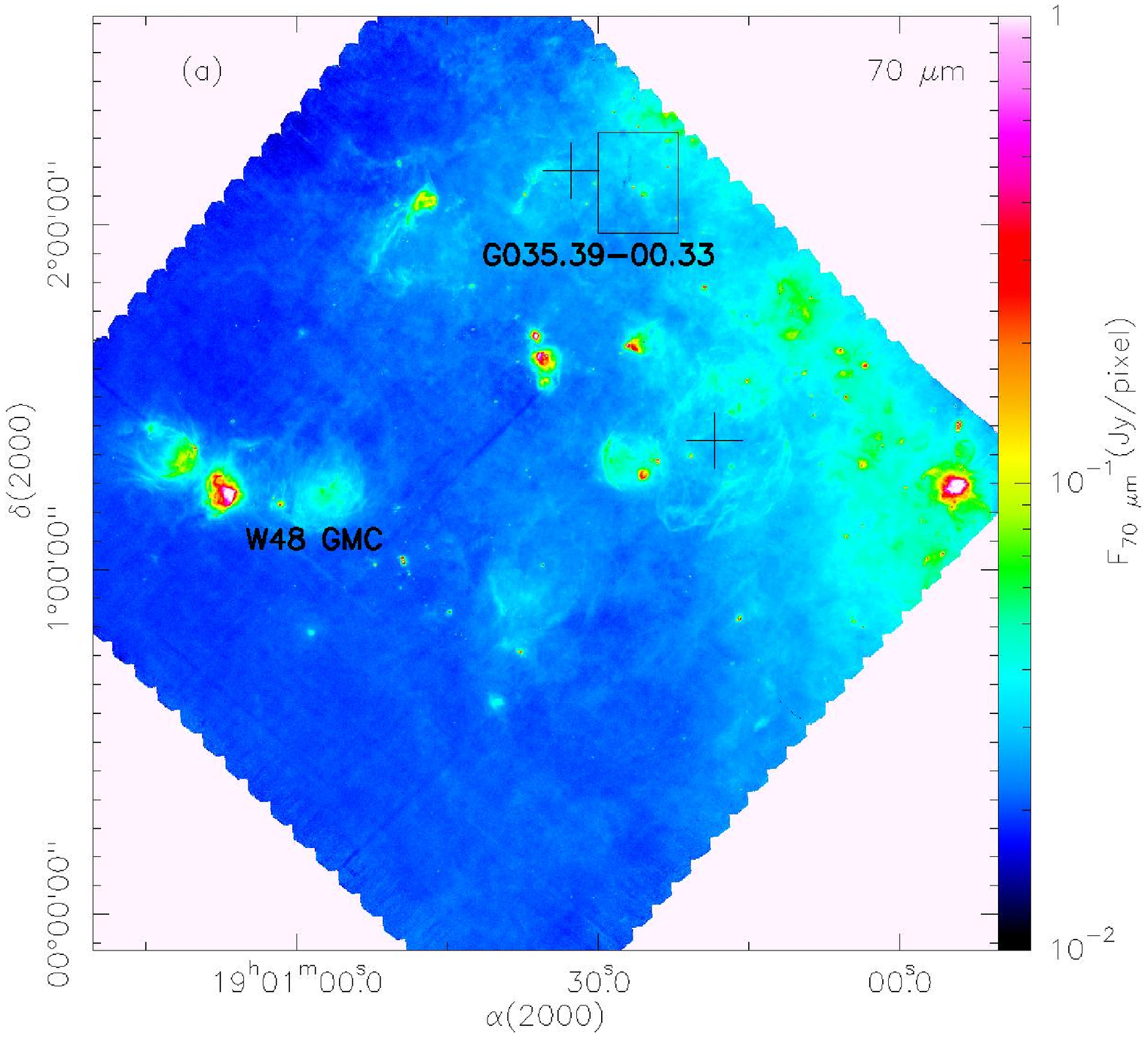}   & 
 \includegraphics[angle=0,width=8.cm]{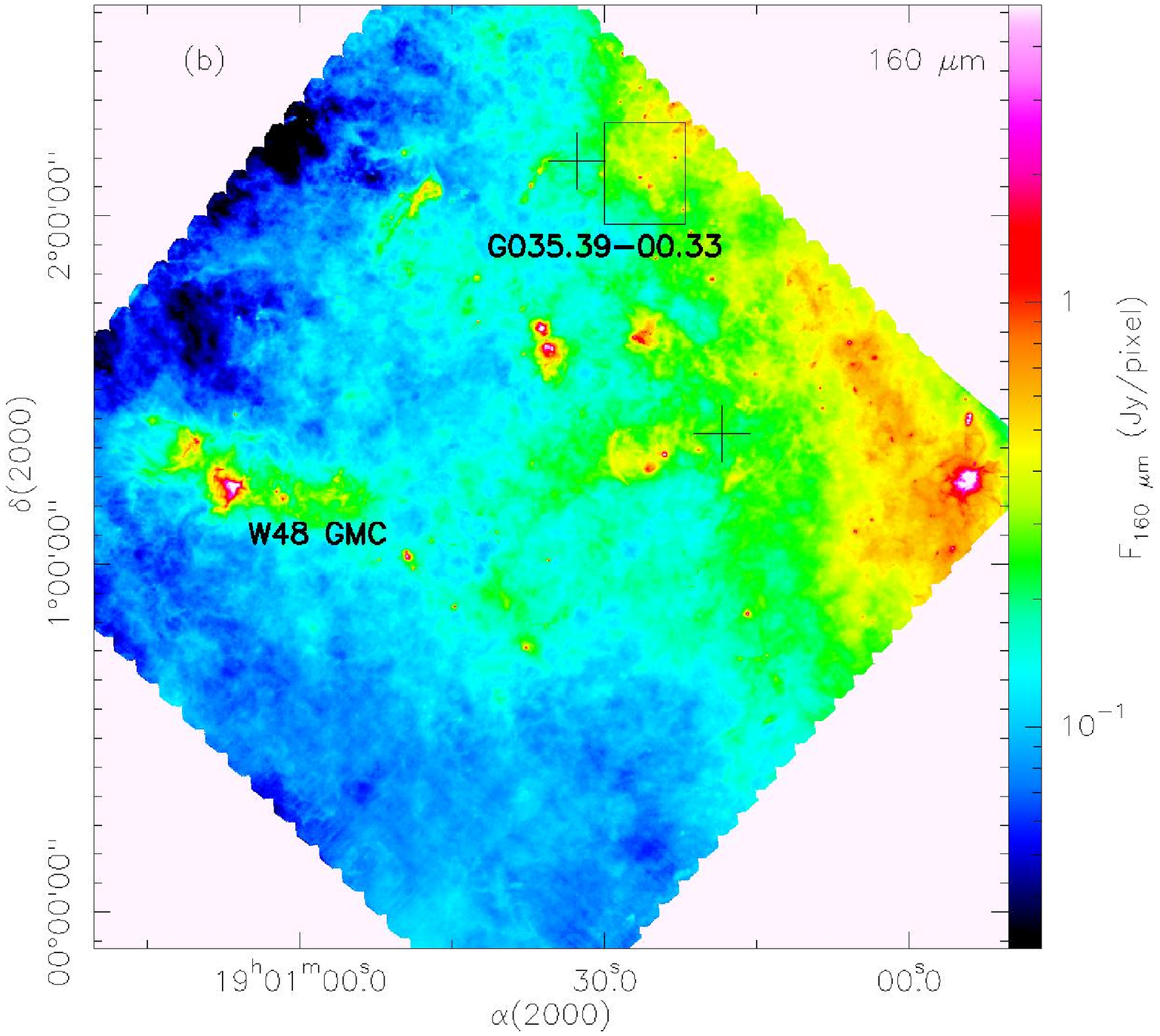} \\
\includegraphics[angle=0,width=8.cm]{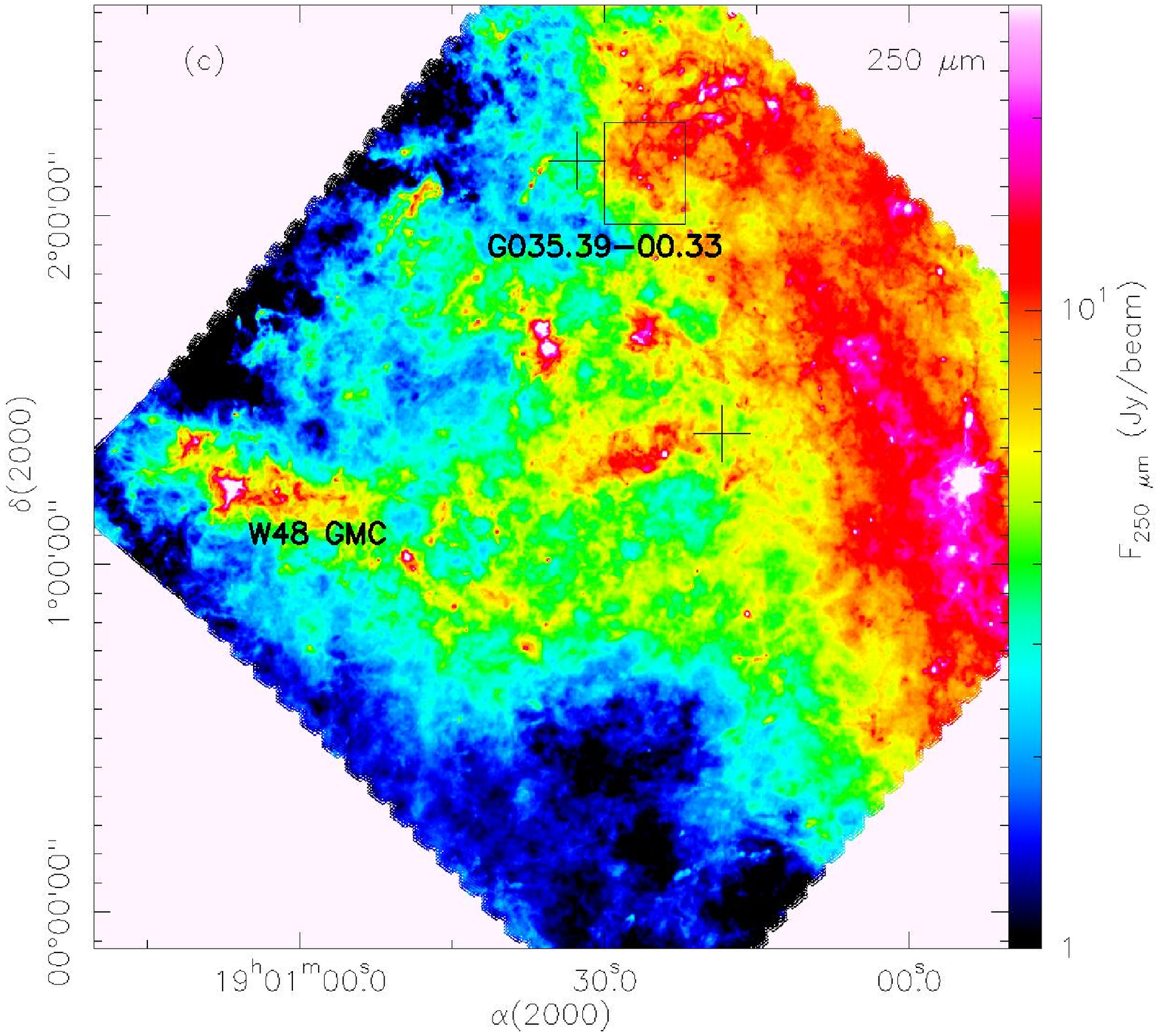}   &  
\includegraphics[angle=0,width=8cm]{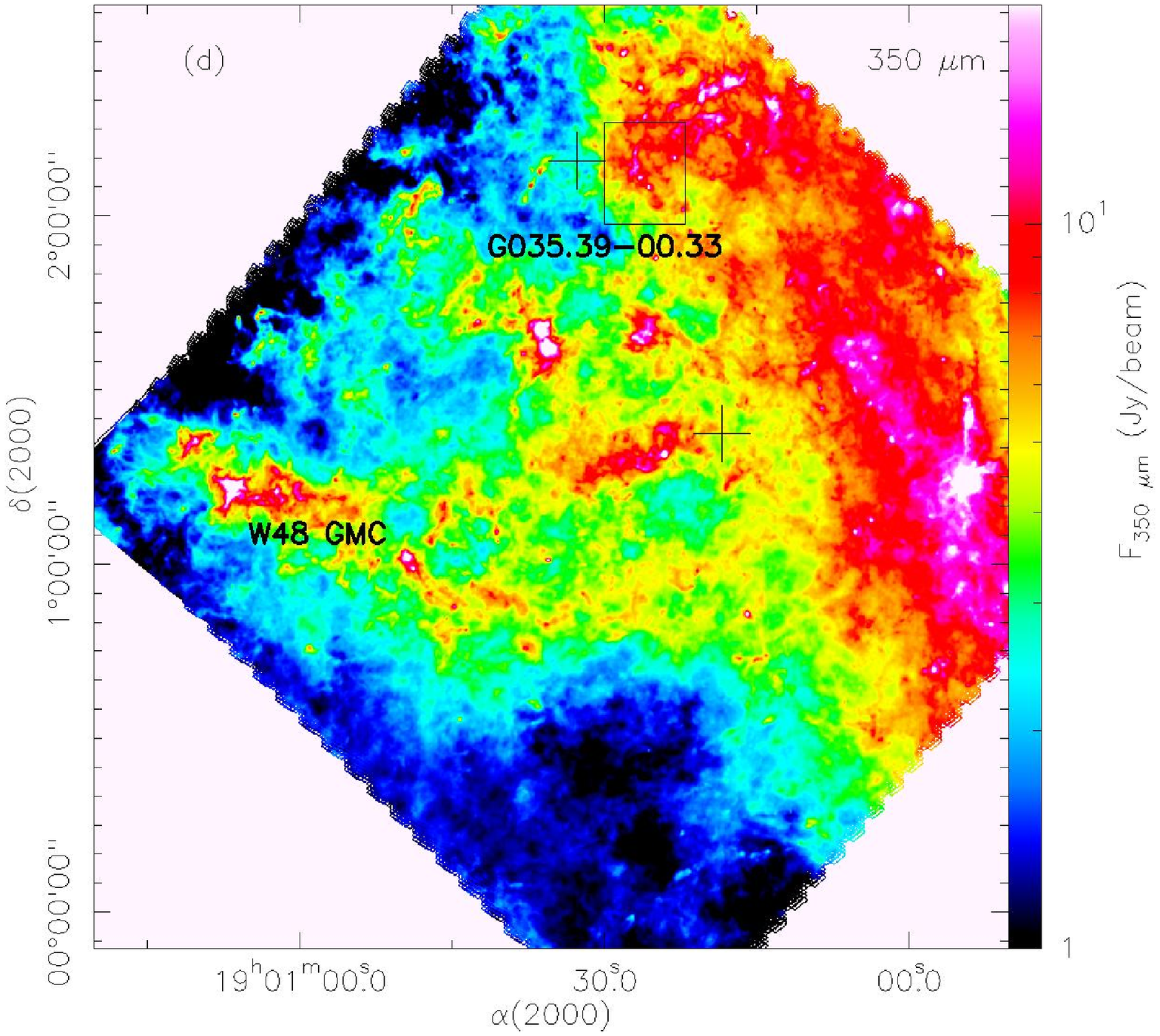} \\
 \includegraphics[angle=0,width=8.cm]{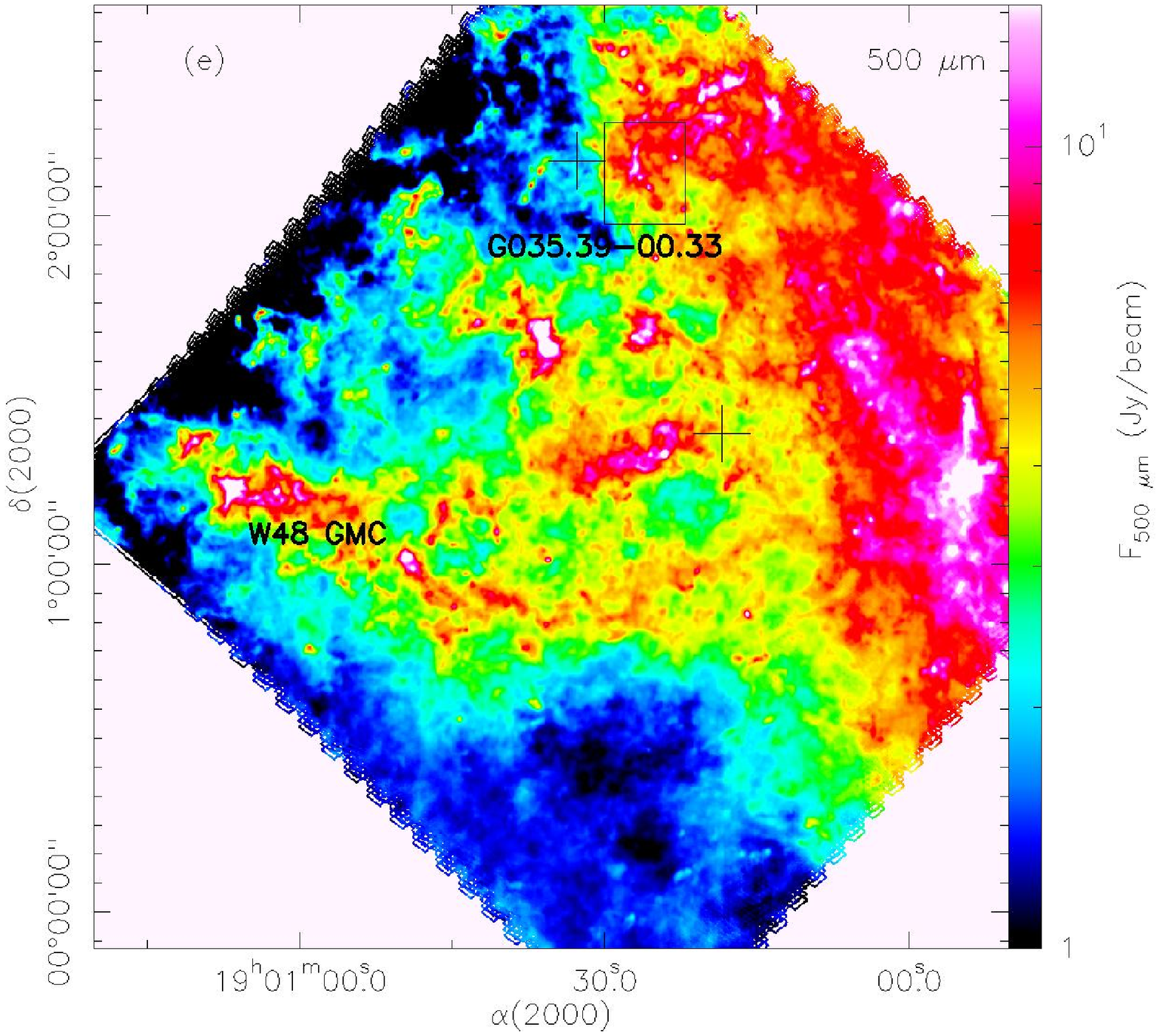}   &  
 \end{array}$
\vskip -0.3cm
\caption{\textit{Herschel} images of W48: {\bf a}) PACS 70~$\micron$ ($HPBW\sim6\arcsec$), {\bf b}) PACS $160~\micron$ ($HPBW\sim12\arcsec$), {\bf c}) SPIRE 250~$\micron$ ($HPBW\sim18\arcsec$), {\bf d}) SPIRE 350~$\micron$ ($HPBW\sim25\arcsec$), and {\bf e}) SPIRE 500 $\micron$ ($HPBW~37\arcsec$). The bright diffuse emission on the right of each panel is from the Galactic plane. Plus signs (+) indicate the location of supernova remnants.
}
\label{fig:imagesall}
\vskip -0.5cm
\end{figure*}

\begin{figure*}[hbtp]
\centering
\includegraphics[angle=0,width=18.cm]{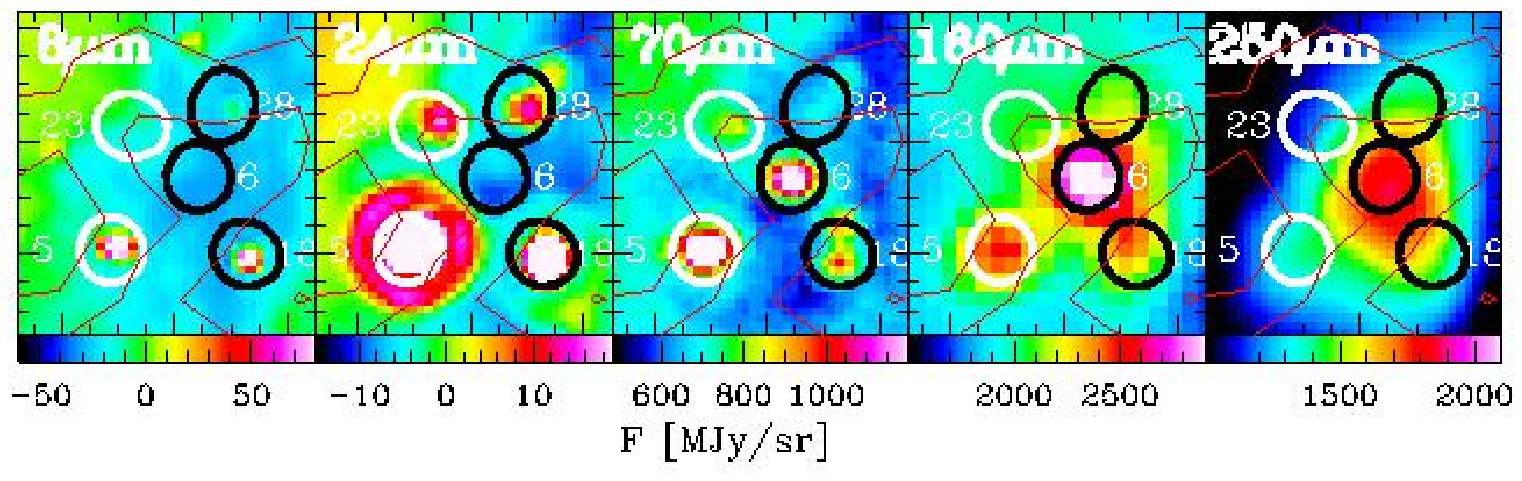} \\
\includegraphics[angle=0,width=18.cm]{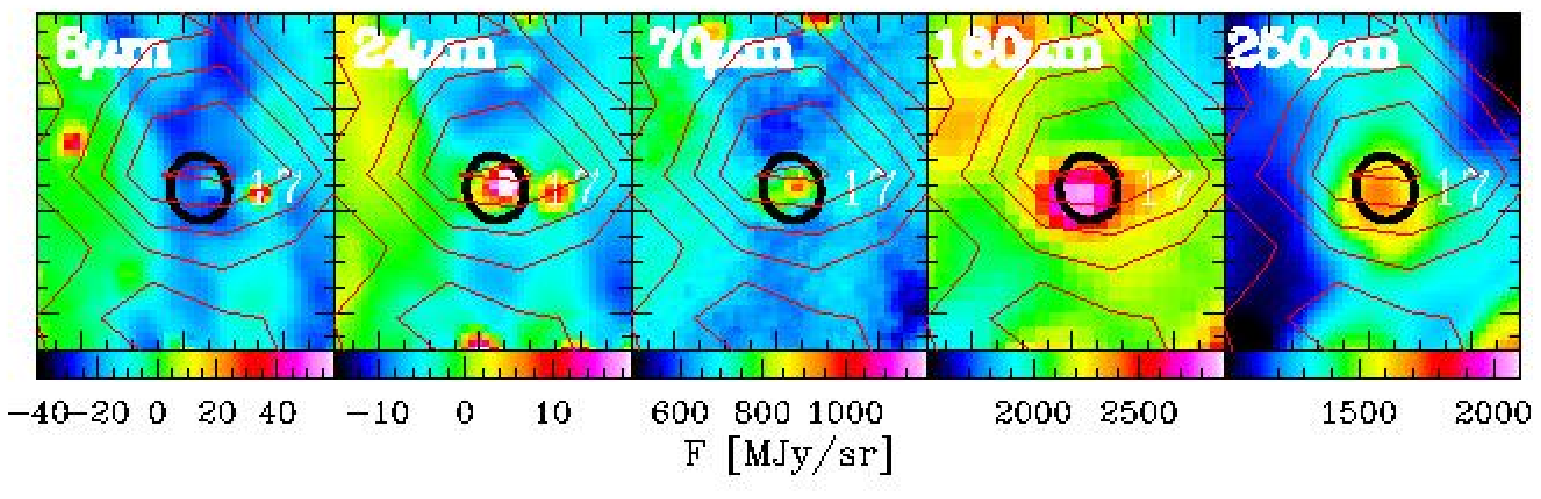} 
\caption{Zooms towards MDC \#6 ({\bf upper rows})  and MDC \#17 ({\bf lower rows}) at 8 - 250~$\micron$. Cloud fragments with mass $<$$20~\msun$ are marked as white and MDCs as black ellipses. SiO contours are the same as in Fig.~\ref{fig:Sio} (from 0.01 to 0.05~K$\,\kms$ by 0.01~K$\,\kms$ from \citealt{jimenez-serra10}).}
\label{fig:sionorthsouth}
\vskip 0.5cm
\end{figure*}

\begin{figure*}[!h]
\centering
$
\begin{array}{ccc}
\includegraphics[angle=0,width=18.cm]{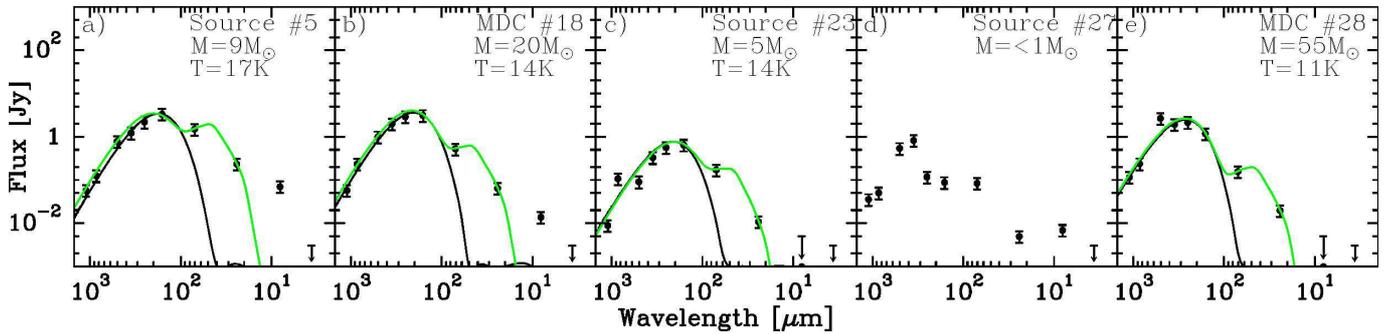} 
\end{array}
$
\vskip -0.3cm
\caption{
SEDs built from \textit{Herschel} and other wavelengths for sources lying towards the SiO peaks (except for those shown in Fig.~\ref{fig:SEDex}). The curves are grey-body models fitted for data at wavelengths $\ge$~160~$\micron$. The one-temperature grey-body fit (black curve) is consistent with two-temperature grey-body fits (green curve). Error bars correspond to 30\% of the integrated fluxes. No model can reasonably fit source \#27.
}
\label{fig:SEDex2}
\vskip -0.3cm
\end{figure*}

\end{document}